\documentclass[aps,prl,twocolumn,preprintnumbers,10pt,showpacs]{revtex4-1}
\usepackage{epsfig,psfrag,amsmath,amssymb,lmodern}
\usepackage[normalem]{ulem}
\usepackage{color}
\usepackage[applemac]{inputenc}

\usepackage[T2A,T1]{fontenc}
\usepackage{tikz}
\usetikzlibrary{decorations.markings}
\tikzset{->-/.style={decoration={
  markings,
  mark=at position #1 with {\arrow{>}}},postaction={decorate}}}
 \tikzset{-<-/.style={decoration={
  markings,
  mark=at position #1 with {\arrow{<}}},postaction={decorate}}}

\usepackage{bm}

\allowdisplaybreaks[4]

\def \nn {\nonumber}

\newcommand \ket [1] {|{#1}\rangle}
\newcommand \bra [1] {\langle {#1}|}

\newcommand{\cM}{{\cal M}} 

\newcommand{\cN}{{\cal N}}

\newcommand{\cG}{{\cal G}}
\newcommand{\cO}{{\cal O}}

\newcommand{\nt}{\notag\\}

\newcommand{\cL}{{\cal L}}
\newcommand{\qq}{\,, \qquad }

\definecolor{darkred}{rgb}{0.5,0.0,0.0}
\definecolor{darkblue}{rgb}{0.0,0.0,0.9}
\definecolor{darkerblue}{rgb}{0.0,0.0,0.5}
\definecolor{darkgreen}{rgb}{0.0,0.5,0.0}
\definecolor{black}{rgb}{0.0,0.0,0.0}
\definecolor{brown}{rgb}{0.6,0.4,0.2}
\newcommand{\red}{\color{darkred}}
\newcommand{\blue}{\color{darkerblue}}

\def \be  {\begin{equation}}
\def \ee  {\end{equation}}
\def \ba  {\begin{eqnarray}}
\def \ea  {\end{eqnarray}}
\def \baa {\begin{eqnarray*}}
\def \eaa {\end{eqnarray*}}
\def \bb  {\begin {thebibliography} }
\def \eb  {\end{thebibliography}}

\makeatletter
\renewcommand*\env@matrix[1][*\c@MaxMatrixCols c]{%
  \hskip -\arraycolsep
  \let\@ifnextchar\new@ifnextchar
  \array{#1}}
\makeatother

\def\N {{N}}

\begin{document}
\title{
\begin{flushright} ${}$\\[-40pt] $\scriptstyle \rm  LAPTH-004/24 $ \\[0pt]
\end{flushright}
The Collinear Limit of the Four-Point  Energy Correlator\\ in  $\mathcal{N}=4$ Super Yang-Mills Theory
}

\author{ 
Dmitry Chicherin,$^{a}$ Ian Moult,$^{b}$ Emery Sokatchev,$^{a}$ Kai Yan,$^{c,d}$  Yunyue Zhu$^{c}$
}
\affiliation{
$^a$ LAPTh, Universite Savoie Mont Blanc, CNRS, B.P.110,  F-74941 Annecy-le-Vieux, France \\
$^b$ Department of Physics, Yale University, New Haven, CT 06511, USA  \\
$^c$ School of Physics and Astronomy, Shanghai Jiao Tong University, Shanghai 200240, China \\
$^d$ Key Laboratory for Particle Astrophysics and Cosmology (MOE), Shanghai 200240, China
}

\begin{abstract}
We present a compact formula, expressed in terms of classical polylogarithms up to weight three, for the leading order four-point energy correlator in maximally supersymmetric Yang-Mills theory, in the limit where the four detectors are collinear. This formula is derived by combining a simplified, manifestly dual conformal invariant form of the $1\to 4$ splitting function obtained from the square of the tree-level five-particle form factor of stress-tensor multiplet operators, with a novel integration-by-parts algorithm operating directly on Feynman parameter integrals.  Our results provide valuable data for exploring the structure of physical observables in perturbation theory, and for calculations of jet substructure observables in quantum chromodynamics.
\end{abstract}

\maketitle

\subsection{Introduction}  
Explicit calculations of observables in quantum field theory (QFT) have played a crucial role in uncovering hidden simplicity and structure, and in providing theoretical data for the development of new computational approaches. While a large amount of data exists for perturbative scattering amplitudes, less attention has been focused on physical cross-section level observables relevant for colliders. 
Motivated by the remarkable simplicity hidden in scattering amplitudes, we can hope for more exciting surprises in the study of physical observables. 

%
%

An interesting class of physical observables are $N$-point correlation functions of energy flux \cite{Basham:1977iq,Basham:1978bw,Basham:1978zq,Basham:1979gh}, which we denote $\rm{E^N C}$. The $\rm{E^N C}$ are infrared finite  \cite{Kinoshita:1962ur,Lee:1964is}, exhibit an operator product expansion \cite{Hofman:2008ar,Kologlu:2019mfz,Chang:2020qpj}, and can be directly measured in experiment \cite{OPAL:1990reb,ALEPH:1990vew,L3:1991qlf,SLD:1994yoe,Komiske:2022enw,Chen:2022swd,Lu:2023,Fan:2023,Tamis:2023guc}, making them ideal candidates for explorations in perturbation theory. Although the $\rm{E^N C}$ was computed at strong coupling \cite{Hofman:2008ar} in $\mathcal{N}=4$ super-Yang-Mills (sYM), 
little is known about its  structure at weak coupling. The $\rm{E^2 C}$ was computed at leading-order in the coupling expansion (LO) in quantum chromodynamics (QCD) in the seminal work of \cite{Basham:1978bw}, and more recently at next-to-leading order (NLO) in QCD \cite{Dixon:2018qgp, Luo:2019nig} and NNLO in $\cN=4$ sYM \cite{Belitsky:2013ofa,Henn:2019gkr}. Explorations of higher point correlators were initiated with the calculation of the ${\rm{E^3C}}$ in the triple-collinear limit \cite{Chen:2019bpb}, and for generic angles \cite{Yan:2022cye,Yang:2022tgm}. These calculations provided important data for the light-ray OPE \cite{Chen:2020adz,Chen:2021gdk,Chen:2022jhb,Chang:2022ryc}, and were crucial for developing new collider physics observables \cite{Komiske:2022enw,Chen:2022swd},  motivating the exploration of higher point correlation functions.




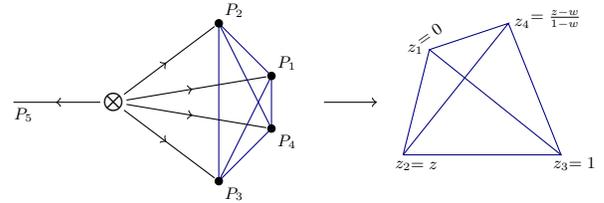
\begin{figure}[t] 
\begin{tikzpicture}[scale=0.70, transform shape]
    \coordinate (A) at (0,1.5);
    \coordinate[label=above right:$P_2$] (B) at (2,3);
    \coordinate[label=above right:$P_1$] (C) at (3,2);
    \coordinate[label=below right:$P_4$] (D) at (3,1);
    \coordinate[label=below right:$P_3$] (E) at (2,0);
    \coordinate[label=below right:$P_5$] (F) at (-2,1.5);
    \draw[->-=0.5, shorten >=2pt, shorten <=5pt] (A) -- (B);
    \draw[->-=0.5, shorten >=2pt, shorten <=5pt] (A) -- (C);
    \draw[->-=0.5, shorten >=2pt, shorten <=5pt] (A) -- (D);
    \draw[->-=0.5, shorten >=2pt, shorten <=5pt] (A) -- (E);
    \draw[-<-=0.5, shorten >=5pt, shorten <=2pt] (F) -- (A);
    \draw[darkerblue] (B) -- (C);
    \draw[darkerblue] (B) -- (D);
    \draw[darkerblue] (B) -- (E);
    \draw[darkerblue] (C) -- (D);
    \draw[darkerblue] (C) -- (E);
    \draw[darkerblue] (D) -- (E);
    \foreach \node in {B,C,D,E} {
        \filldraw (\node) circle (2pt);
    }
    \node at (A) {$\bigotimes$};
    
    \draw[->] (4,1.5) -- (5,1.5);
    
    \coordinate[label=below:$z_2$] (O) at (6-0.5,0.5);
    \coordinate[label=below:$z_3$] (P) at (9-0.5,0.5);
 \coordinate[label=left:$z_1$] (Q) at (6.5-0.5,2.5);
    \coordinate[label=right:$z_4$] (R) at (8-0.5,3);
    \node[rotate=35] at (6.5-0.5, 2.8)    {$=0$};
    \node[] at (6.4-0.5,0.3) {$= z $};
     \node[] at (9.4-0.5,0.35) {$ =1 $};
      \node[] at (8.9-0.5,3.1) {$= \frac{z-w}{1-w} $};
    \draw[darkerblue] (O) -- (P) -- (Q) -- (R) -- (O);
    \draw[darkerblue] (O) -- (Q);
    \draw[darkerblue] (P) -- (R);
\end{tikzpicture}
\caption{The collinear limit of the four-point energy correlator parametrized in terms of momenta, see Eq.~(\ref{2.16}), and complex parameters $(z,\bar z)$ and $(w, \bar w)$, see Eq.~(\ref{eq:zwparam}). 
}
 \label{fig:kinematic}
\end{figure}


A simplified limit of the $\rm{E^N C}$ is the multi-collinear limit, shown in Fig. \ref{fig:kinematic}. In this limit, the $\rm{E^N C}$ becomes independent of the source, and reduces to a function of $2(N-2)$ independent angles. 
This limit is also motivated by jet substructure \cite{Larkoski:2017jix,Asquith:2018igt}, where the $N$ detectors all lie in a single jet \cite{Dixon:2019uzg,Chen:2020vvp,Lee:2022ige,Chen:2023zlx}.  At LO  the collinear limit of the ${\rm E^N C}$ is given as a phase space integral of the tree-level universal $1\to \N$ splitting functions \cite{Amati:1978wx,Amati:1978by,Ellis:1978sf,Catani:1998nv}, $\mathcal{P}^{(0)}_{1\rightarrow \N}$, as
\begin{align}
 {\rm E^N C}\overset{\rm coll.}{=} 
 \hskip-1.0mm  \int_0^1 \hskip-.5mm d x_1 \cdots d x_\N \, \delta(1- \sum_i x_i) \,  (x_1 \cdots x_\N)^2\, \mathcal{P}^{(0)}_{1\rightarrow \N} \notag \,,
\end{align}
where  $x_i$ is the  fraction of energy carried by the particle absorbed by detector $i$. 
Remarkably,  the $N-$point correlators define a class of  manifestly \emph{finite integrals} in $(N-1)-$dimensional projective space, making it desirable to develop methods that directly compute in four dimensions, profiting from the simplifications of \emph{finite} Feynman loop integrals \cite{Caron-Huot:2014lda,Henn:2022vqp,Gambuti:2023eqh}.
This structure is similar to the Feynman-parameter representation of loop integrals, allowing the application of recently developed techniques \cite{Artico:2023bzt,Artico:2023jrc,Britto:2023rig,Chen:2019mqc,Lee:2014tja}. 
Given this richness in both physical and mathematical contexts, it is imperative to develop efficient  techniques for {\it finite} integrals, well-adapted to the computation of the $\rm E^\N C$ for arbitrary $\rm \N$. 


In this \emph{Letter} we obtain  a new result for the LO four-point correlator in $\mathcal{N}=4$ sYM  by squaring a $1\to 5$ form factor to derive a simplified expression for the $1\to 4$ splitting function, which is then 
 integrated using a novel integration-by-parts technique that operates directly in the energy parameter $(x_i)$ space.   


  
\subsection{Super Form Factors and Splitting Functions}






\begin{figure}[t!]
\begin{align*}
\hskip-3mm
\begin{array}{c}\includegraphics[width=3.8cm]{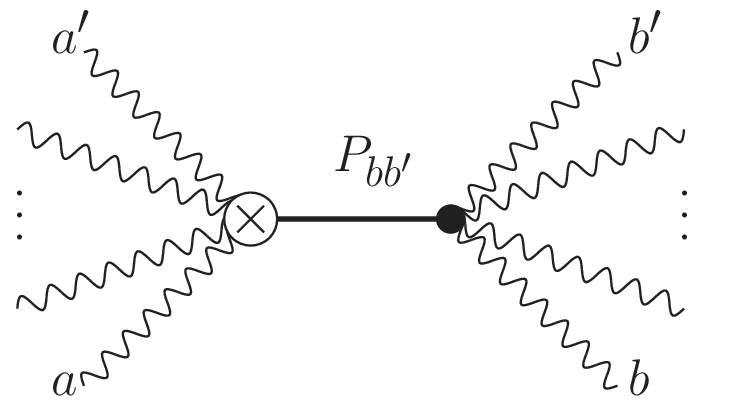}
\end{array}
\begin{array}{c}
\bm{\times}
\end{array}
\begin{array}{c}\includegraphics[width=3.8cm]{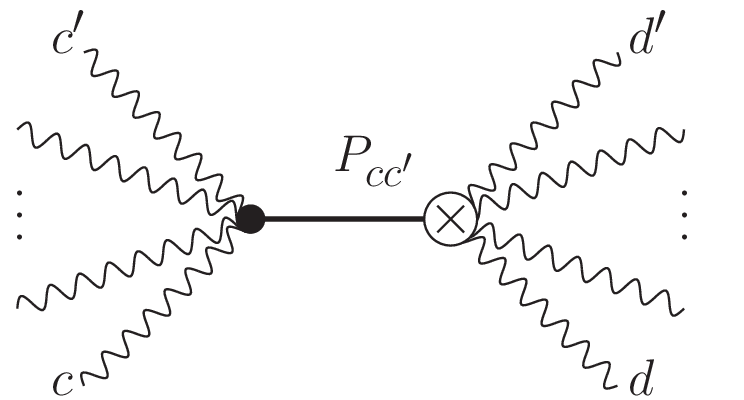}
\end{array} 
\end{align*}
\caption{The product of chiral and antichiral MHV diagrams used to compute the form factor squared.} \label{fig_NMHVxNMHV}
\end{figure}

We begin by deriving a compact form for the tree-level form factors squared, $|F_{N+1}(\cO)|^2=|\bra{0}\cO(q)\ket{p_1,h_1,\ldots,p_{N+1},h_{N+1}}|^2$, for $N=3,4$, where $\cO$ is a member of the  stress-tensor multiplet in planar $\cN=4$ sYM. Here $(p_i,h_i)$ denote massless on-shell momenta and helicities. The particles in the final state (gluons, gluinos and scalars) form on-shell supermultiplets. The stress-tensor multiplet includes, amongst others,  the protected half-BPS operator $\cO_{20'}$, the stress-energy tensor and the Lagrangian $\cL$ of the theory. $\cN=4$ supersymmetry  relates their form factors squared  and summed over all helicities in the on-shell multiplets,  in a  simple way, e.g., $ |F_{N+1}(\cL)|^2 = (q^2)^2 |F_{N+1}(\cO_{20'})|^2$, with $q=\sum_{i=1}^{N+1} p_i$ being the total injected momentum. 


To compute the splitting function $\mathcal{P}^{(0)}_{1\rightarrow \N}$, we consider the multi-collinear limit of $|F_{N+1}|^2$, in which the angles between the first $N$ particles become small simultaneously (see Fig.~\ref{fig:kinematic}). 
The on-shell particle momenta $p^\mu$  
 have the lightcone components 
 \begin{align}\label{2.16}
& p^+_{i} = x_i   \qq  p^-_{i}=x_i |z_i|^2 \,, \qquad  p^\perp_{i } = x_i z_i  \,,
\end{align}
where the energy fractions $x_i ={2p^0_i}/{q^0}$  satisfy $\sum_1^{N+1} x_i = 2$. In the limit the complex parameters $z_i $, for $i=1,\ldots,N$, go to zero at the same rate. We also have $x_{N+1}=1$ and we set $z_{N+1}=\infty$ by a Lorentz transformation.  The  Mandelstam variables simplify to
\begin{align*}
&s_{i_1 i_2 \ldots i_k N+1} =x_{i_1}+x_{i_2}+\ldots+x_{i_k} \equiv x_{i_1 i_2 \ldots i_k}\,,\\
& {\blue s_{i_1 i_2 \ldots i_k}} = \sum_{i<j}   
 x_i x_j |z_i-z_j|^2 , \ \,\,   i,j = 1, \ldots, N .
\end{align*}

For the case $N=4$,  $\cN=4$ supersymmetry implies that we need only the square of the sum of the MHV and NMHV $F_5(\cL)$.  Tree-level form factors in $\cN=4$ sYM are computed efficiently by the so-called MHV rules \cite{Cachazo:2004kj,Brandhuber:2011tv}. We consider the planar limit, $N_c \to \infty$, and specify a single color orientation. The MHV matrix element squared is 
$\mathbb{F}\overline{\mathbb{F}}^{\rm MHV}_{N+1} = N_c^{N+1}  {(q^2)^4}/({s_{12}s_{23} \ldots s_{1(N+1)}})$.
The NMHV form factor consists of a sum of diagrams shown on the left panel in Fig.~\ref{fig_NMHVxNMHV} (the right panel corresponds to the conjugate $\overline{\rm NMHV}$ form factor). They are labeled anti-clockwise  by $b'>b \in [1,N+1]$ denoting the set of particles coming out of the interaction vertex. The complementary set with $a'=b'+1$ and $a=b-1$ consists of particles connected to the (non-polynomial) Lagrangian operator, denoted by a cross. The momentum $P_{bb'}=\sum_{i=b}^{b'} p_i$ flows from the operator to the interaction vertex.  In the multi-collinear limit, if the non-collinear leg $N+1$ is on the interaction side, the contribution of the graph to the highest order pole vanishes.    Finally, the matrix element squared $\mathbb{F}\overline{\mathbb{F}}_{N+1}$ is obtained by multiplying the sum of all diagrams of the same color orientation by its conjugate  and summing over the helicities of the final states. We find that the ratio between the  $(N+1)-$point  $\rm{NMHV} \times \overline{\rm{NMHV}} $ and $\rm{MHV} \times \overline{MHV}$ form factors  in the  multi-collinear limit $1||2\dots ||N$ is given by
\begin{widetext}
\begin{align}\label{eq3}
 & \frac{\mathbb{F} \overline{\mathbb{F}}^{\rm NMHV}_{N+1}}{\mathbb{F}\overline{\mathbb{F}}^{\rm MHV}_{N+1}} \overset{\rm coll.}{=}  \sum_{\{b,b'\};\{c,c'\}}  \Bigg(\sum_{i=b}^{b'} \sum_{j=c}^{c'} \sum_{k=e}^{e'}  x_i x_j x_k z_{i k } \bar z_{k j} \Bigg)^4  \times
  \frac{z_{b-1\, b}z_{b'\, b'+1} }
{ {\blue s_{b\ldots b'}}  K_{b-1,b'}  K_{b,b'}  L_{b,b'} L_{b,b'+1} }  
\frac{ \bar z_{ c-1\,  c} \bar z_{ c'\, c'+1} }
{ {\blue s_{c \ldots c'}} \overline{K}_{ c-1,c' }  \overline{K}_{c,c' }  \overline{L}_{c,c' }   \overline{L}_{  c,c'+1 } }\,, \nt
&  z_{ij}=z_i-z_j, \quad K_{ab} \equiv  \sum_{i=a}^b z_{a i} x_{i}, \quad  L_{a b} \equiv \sum_{i=a}^b z_{bi } x_i , \quad  1 \leq b  < b' \leq  N\,, \; 1 \leq c  <c' \leq  N, \;  [e,e'] = [b,b'] \cap [c,c']  \,.
\end{align}
\end{widetext}
The unphysical poles in Eq.~(\ref{eq3}) cancel out in the sum of all terms. 
$\cN=4$  supersymmetry implies a helicity decomposition of the matrix elements squared, e.g.,
$\mathbb{F}\overline{\mathbb{F}}_{5} = 2\mathbb{F}\overline{\mathbb{F}}_{5}^{\rm MHV} +2\mathbb{F}\overline{\mathbb{F}}_{5}^{\rm NMHV}$.
The splitting function is obtained from the multi-collinear limit of  $\mathbb{F}\overline{\mathbb{F}}_{N+1}$ by symmetrizing over all permutations of the collinear particles, and by dividing by the two-particle matrix element squared to cancels the dependence on the operator ${\cal L}$,
\begin{widetext}
\begin{align}\label{splitting}
{\cal P}^{(0)}_{1 \to N} =  {N_c^{N-1} \over (x_1 \ldots x_N)^2 |z_{12} \ldots z_{N-1 N}|^2}\,  \cG_N + \text{perm}(1,2,\ldots,N) \quad \text{where} \quad \cG_{N} := \lim_{1||2\dots ||N} { \mathbb{F}\overline{\mathbb{F}}_{N+1} \over \mathbb{F}\overline{\mathbb{F}}^{\rm MHV}_{N+1}}\ .
\end{align}
\end{widetext}

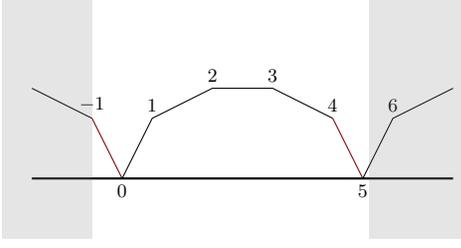
\begin{figure}[t]
\begin{tikzpicture}[scale=0.80, transform shape]
    \coordinate[label=above:$-1$] (P-1) at (1,0);
    \coordinate[label=below:$0$] (P0) at (1.5,-1);
    \coordinate[label=above:$1$] (P1) at (2,0);
    \coordinate[label=above:$2$] (P2) at (3,0.5);
    \coordinate[label=above:$3$] (P3) at (4,0.5);
    \coordinate[label=above:$4$] (P4) at (5,0);
    \coordinate[label=below:$5$] (P5) at (5.5,-1);
    \coordinate[label=above:$6$] (P6) at (6,0);
    \draw (0,0.5) -- (P-1); 
    \draw[darkred] (P-1) -- (P0);
    \draw (P0) -- (P1) -- (P2) -- (P3) -- (P4);
    \draw[darkred] (P4) -- (P5);
    \draw (P5) -- (P6) -- (7,0.5);
    \draw [thick] (0,-1) -- (P0) -- (P5) -- (7,-1);
    \fill[gray,opacity=0.2] (-0.5,-2) rectangle (1.0,2);
    \fill[gray,opacity=0.2] (5.6,-2) rectangle (7.2,2);
\end{tikzpicture}
\caption{\small Dual coordinates for the five-particle form factor. The points lie on an infinite contour with a period equal to the injected momentum $q$.
}
\label{fig:dual_FF}
\end{figure}

Our general result in Eq.~(\ref{eq3}) reproduces that for $N=3$  in \cite{Chen:2019bpb},  but in a way which makes one of the remarkable features of amplitudes \cite{Drummond:2006rz,Drummond:2007aua,Drummond:2008vq} and form factors \cite{Alday:2007he,Maldacena:2010kp,Bork:2014eqa, Bianchi:2018rrj,Ben-Israel:2018ckc} in $\cN=4$ sYM manifest, namely the dual conformal symmetry acting linearly on the particle momenta. To this end we introduce dual coordinates $y_i$ by the rule $p_i=y_i-y_{i-1}$.  In the amplitude case with vanishing total momentum $q=0$ the dual points form a closed contour. For the form factor $q\neq0$, and the dual points lie on an infinite contour with period $q$, such that $y_{ij} = y_{i+N+1, j+N+1}$ and $y_{i+N+1,i}\equiv q$, where $y_{i,j} \equiv y_i - y_j$. 
 This is illustrated in Fig.~\ref{fig:dual_FF} for the relevant case of the five-particle form factor. To reproduce the result of \cite{Chen:2019bpb}, we choose the points $y_0,\ldots,y_5$ on the contour  and write the result in terms of dual conformal cross-ratios $(a,b,c,d)={y^2_{ab} y^2_{cd}}/({y^2_{ac}y^2_{bd}})$,
\begin{align*}
\hspace{-0.25cm}
\hskip-0.5mm
 \frac{ \mathbb{F}\overline{\mathbb{F}}^{\rm NMHV}_{4}}{\mathbb{F}\overline{\mathbb{F}}^{\rm MHV}_{4}} \overset{\rm coll.}{=}   ({\red -1},1,2, {\red4}) + ({\red -1},3,2,0) + (3,1,0, {\red 4}) \,.
\end{align*}
A new result reported here is the splitting function for $N=4$ in $\mathcal{N}=4$ sYM. It is given by a very compact dual conformal function of the points $y_i$ with $i=-1,0,\ldots,5$:
\begin{widetext}
\begin{align*}
\frac{\mathbb{F} \overline{\mathbb{F}}^{\rm NMHV}_{5}}{\mathbb{F}\overline{\mathbb{F}}^{\rm MHV}_{5}}  \overset{\rm coll.}{=}   &-1+  ({\red -1},1,  2, {\red 5})  +   ( {\red -1}, 2, 3,  {\red 5} )   +
({\red -1}, 4,  3,0) +  (4, 1, 0 ,{\red 5}) 
  + ({\red -1}, 3,  2,0) +  (4, 2, 1 ,{\red 5})   +(0,4, 3, 1) \nt
  &+ (0,4, 3, 1)  ({\red -1},1, 3,{\red 5}) 
+   ({\red -1}, 4, 3,1) ( 3,1, 0, {\red 5})  + ({\red -1}, 4, 2, 0)( 0,2,3 , {\red 5})\nt 
& +   ({\red -1}, 1, 2, 4)( 4,2,0 , {\red 5}) 
+   ({\red -1}, 3, 2, 0)( 4,2,0 , {\red 5})  +    ({\red -1}, 4, 2, 0)( 4,2, 1 , {\red 5})  + ({\red -1}, 4, 3,0) ({\red -1}, 1,2,4) \nt 
&  + (4,1,0, {\red 5})(  0,2,3,{\red 5})  
+   ({\red -1}, 4, 3,1) ({\red -1}, 4,2,0) +  (3,1,0, {\red 5})(  4,2,0,{\red 5})  \nt 
& + ({\red -1}, 4, 3, 1)( {\red -1}, 1,2 , {\red 5})  +   ( 3,1,0, {\red{5}})( {\red -1}, 2,3 , {\red 5})
+  ({\red -1}, 1,2,4)( {\red -1}, 1,3 , {\red 5}) +  (0,2,3, {\red{5}})( {\red -1}, 1,3 , {\red 5}) \,. 
\end{align*}
\end{widetext}


\subsection{Phase-space Integration from Integration-by-Parts}


To compute the $\rm E^\N C$ in the collinear limit we invent an automated method exploiting simplifications from integer-dimensional integration-by-parts (IBPs) identities. Further developing algorithms from \cite{Henn:2022vqp}, we show how to find a basis of master integrals which is both free from  divergences and avoids higher-degree poles,  and illustrate this algorithm for the specific case of the $\rm E^4 C$.



\medskip
\noindent \textbf{1. Family Definitions:} 
%
%
The splitting function  exhibits a uniform scaling weight in the energy fractions, allowing them to be treated  as projective coordinates $[x_1: \cdots: x_\N] \in {\rm P}_{\N-1}(\mathbb{R}_{+})$,  parametrizing the $\N-$body collinear phase space.  This yields a master formula for the ${\rm E^\N C}$ in the collinear limit,
\begin{align}\label{NpointDef}
 {\rm E^\N C}  & \overset{\text{coll.}}{=} 
 \frac{1}{|z_{12} \cdots z_{\tiny{\N-1\N}}|^2}   \hskip-0.5mm
\int  \hskip-0.5mm \frac{d^\N x}{\text{GL}(1)} (x_1+\cdots + x_\N)^{-\N} \cG_{\N}    \nn\\
 & \; +\;  \text{perm}\,(z_1,\cdots,z_\N)\,,  
\end{align}
%
%
%
%
where  $\cG_{N}$ is defined in Eq.~(\ref{splitting}) and we introduced  the short-hand notation 
\begin{align}
\int  \frac{d^\N x}{\text{GL}(1)}  \,  \equiv \int_0^\infty  d x_1 \cdots d x_\N  \,    \delta (1- \sum_{i} x_i )\,.   \nn
\end{align} 


Investigating the integrand topologies,  we observe that the integrals for the ${\rm E^4 C}$ belong to a family   $A_{1,1,1,1,1,1,1,0,0,0}$ defined as 
\begin{align} \label{defFamilyA}
& A_{a_1, \cdots, a_{10} } \equiv \int   \frac{d^4 x}{\text{GL}(1)}  \,  \frac{1}{\prod_i  D_i^{a_i} }\,,  \\
 & D_1 = {\blue s_{1234}},  \quad D_2 = {\blue s_{123}}, \quad D_{3} ={ \blue s_{234}} ,  \quad   \nn \\
 & D_4 = x_{1234}, \quad  D_5 = x_{234} , \quad D_6 = x_{123} , \quad D_7 = x_{34} , \quad \nn \\
 &  D_8 = s_{12},  \quad  D_9 = s_{23}, \quad D_{10}= s_{34} \,, \nn 
\end{align}
with multi-particle poles only in  $D_{1-7}$. 
The integrals for the ${\rm E^3 C}$ are given by the simpler family  $B_{1,1,1,1,0,0}$, 
\begin{align} \label{defFamilyB}
& B_{a_1, \cdots, a_{6} } \equiv  \int \frac{d^3 x}{\text{GL}(1)}  
\frac{   s_{12}^{-a_5}  \, s_{23}^{-a_6}  }{ {\blue s_{123}}^{a_1} \, x_{12}^{a_2} \, x_{23}^{a_3} \,  x_{123}^{a_4}  }\,. 
\end{align}

When computing the energy correlators, we wish to work with integrands satisfying the following conditions:

\textbf{(a)} Homogeneity:  The integrand has an overall scaling dimension $(-\N)$ under the transformation  $(x_1, \cdots, x_\N) \rightarrow ( \kappa\,  x_1,  \cdots  , \kappa\,  x_\N)$. 

\textbf{(b)} Finiteness: The integral is free from IR divergences as any subset of energy variables go to zero.   Given condition \textbf{(a)},  condition \textbf{(b)} is equivalent to the UV power-counting behavior  
$   \prod_i D_i^{-a_i} \sim O (\kappa^{-1-|S|}) $  as  $S \rightarrow \kappa \, S\,,  \kappa \rightarrow \infty\,,  \;   \forall\, S \subset \{x_1, \cdots , x_\N \} \,.  $ 
%
%

Such integrals fulfill the same criteria as the  `admissible integrals' in \cite{Henn:2022vqp}, for which four-dimensional IBP methods are particularly well suited. One can reduce the integrals with poles of  degree up to $N$ onto a basis of master integrals of lower complexity. The latter can be evaluated with the \textsc{Maple} program HyperInt \cite{Panzer:2014caa}.

Our IBP algorithm operates directly on the projective coordinates $[x_1: \cdots: x_\N]$. Integrals defined without a dimensional regulator exhibit several distinctive features. First, there exist integrand-level partial fractioning identities  among the integrals with different propagator indices. 
It is advantageous to 
 set up partial-fractioning identities among the (finite) integrands  
\begin{align}\label{defsimpleFamilyA}
& A_{a_1, \cdots, a_7; \; q_1, \cdots, q_4}  \equiv   \int \frac{d^4 x}{\text{GL}(1)}  \, 
 \frac{ x_1^{-q_1}  x_2^{-q_2} x_3^{-q_3}  x_{4}^{-q_4} }{ D_1^{a_1} D_2^{a_2} \cdots D_7^{a_7}} \,,
\end{align}
where we demand that $a_i \geq 0, q_i \leq 0, \forall\, i$, and place an upper bound on the total numerator degree.  This allows us to
set up a  simple basis of finite integrands with $x_i-$monomials in the numerator. 
Second, boundary terms are generated in the IBP relations on the surface of the integration domain: $x_i = 0, \, i \in \{1,2, 3, 4 \} $.
  These boundary terms are given by  lower-point finite integrals over coordinates on the codimension-1 boundary $[x_{j}:x_k: x_l] \equiv [x_1: \ldots \hat{x}_i \ldots: x_4]  $,
\begin{align}\label{defsimpleFamilyB}
& B^{[jkl]}_{a_1,a_2, a_3, a_4; \; q_1, q_2, q_3}  \equiv   \int \frac{d^3 x}{\text{GL}(1)}  \, 
 \frac{ x_j^{-q_1}  x_k^{-q_2} x_l^{-q_3}  }{ {\blue s_{jkl}}^{a_1} \, x_{jk}^{a_2} \, x_{kl}^{a_3}\,  x_{jkl}^{a_4}} \,, 
\end{align} 
which are equivalent to the three-point energy integral family in Eq.~(\ref{defFamilyB}) upon setting $(j,k,l)=(1,2,3)$. Hence our algorithm must proceed iteratively, by generating the relevant lower-point integrals and recycling them for the reduction of the higher-point integral families. 

 \medskip
\noindent \textbf{2. IBP Algorithm:} A generic IBP relation in projective space has the form
\begin{align}\label{FinIBPs}
&\int \frac{d^N x}{{\rm GL}(1)} \,  O_i \circ  f =
- \int \frac{d^{N-1} x}{{\rm GL}(1)}  \,  v  \circ f  \bigg|_{x_ i=0}\,,   \\
&\quad  v =  \prod_{k} x_k^{-q_k}\,,  \quad  f =  \frac{1}{ \prod_{j} D_j^{a_j} }  \,, \nn  
\end{align}
where $O_i = \frac{\partial}{ \partial x_i} \, v\,,\,  i = 1, \cdots, \N $, are differential operators acting on projective coordinates, and
we demand  $a_j \geq 0, \, q_k \leq 0,\,  \forall\, j,k$. 
The validity of the IBP relation requires condition \textbf{(a)} and \textbf{(b)} both fulfilled on the left-hand side of Eq.~\eqref{FinIBPs}. 
Hence we use individual scaling dimensions to impose a well-defined power counting on the IBP differential operator $O_i$ in each integration region.
 This means that under the scaling of any subset of integration variables $ S \rightarrow \kappa \, S$\,,
\begin{align} 
O_i \rightarrow \kappa^{ \beta_S } O_i  \,,  \;  \forall \,  S  \subset \{ x_1, x_2, x_3 ,x_4 \}\,,  \nn 
\end{align} 
where the scaling dimension is given by 
$\beta_S = -\sum_{k \in S} q_k - \delta_{i \in S}\,.$
Given a system of operators graded by individual scaling dimensions, we derive finite IBPs in the following steps: 

({\sl i}) \emph{Seeding}: The seeding process is optimized by an integrand reduction procedure. In a given sector we pair potential vectors $v$ with an integrand $f$, and impose condition \textbf{(a)} and \textbf{(b)} on $O_i \circ f$.  
 We generate partial fractioning identities on finite fields for $v \circ f $, organized in \emph{block triangular form} \cite{Liu:2018dmc,Guan:2019bcx}, and solve for an integrand  basis for IBP generation.


({\sl ii})  \emph{IBP generation}: Applying the differential operators to the integrand, we generate IBP relations $\cM_\N^a$ in the form of Eq.~\eqref{FinIBPs}, where the left-hand side is  expanded over  integrals in the $\N-$point family  and the right-hand side defines the boundary integral families. Each integral that appears in  $\cM_\N^a$  is finite by construction.

({\sl iii})  \emph{Integrand reduction}: $\cM_\N^a$ encloses the complete set of $N-$point finite integrals. 
Among them we generate all partial fractioning identities $\cM_\N^b$ following the sector-by-sector procedure described in step ({\sl i}).

({\sl iv})  \emph{Reducing boundary integrals}:  
When possible, we integrate out the boundary integrals generated on the right-hand-side of $\cM_\N^a$. 
Otherwise, we repeat our procedure on the $(\N-1)$-point integral family.

({\sl v}) \emph{Integral reduction for the full family}: 
We carry out the reduction of $\cM_{\N}^a$ together with $\cM_{\N}^b$ iteratively, starting from $\N=2$. 

In the case of the $\rm{E^4C}$, our algorithm generates $3 \times 10^3$ finite integrals in the form of Eq.~\eqref{defsimpleFamilyA} with total numerator degree $p<7$.  To reduce the target integrals in Eq.~\eqref{NpointDef}, the reduction was carried out for $30^4$ iterations with different kinematic values over finite fields, allowing functional reconstruction powered by \cite{Peraro:2016wsq,Peraro:2019svx}.





\medskip

\noindent \textbf{3. Master Integrals and Symbol Alphabets:}  For the four-point integral family in Eq.~(\ref{defFamilyA}), after performing the integrand reduction, and discarding empty topologies that contain no finite integrals, 
the four-point integrals 
are mapped onto  six sub-topologies, plus their images under a \emph{reflection} symmetry which flips the detector orientation : $1 \leftrightarrow 4, 2 \leftrightarrow 3$. 
The six topologies are  divided into the following three categories: 

Type-I  (one  $3-$particle cut) : \,   ({\blue 2},4,5,6)  \,  ({\blue 2},4,6,7)\,,

Type-II  ($4-$  and  $3-$particle cut) : \, ({\blue{1}},{\blue{2}},4,5) \,  ({\blue 1},{\blue 2},4,7)\,,

Type-III  (two  $3-$particle cuts):\,   ({\blue 2},{\blue 3},4,5)  \,  ({\blue 2},{\blue 3},5,6), 

\noindent where each topology is labeled by the location of its maximal cuts. 
After IBP reduction, we find  28 four-point master integrals $A_{1-28}$, 14 three-point boundary integrals $B_{1-14}$ and one constant function $C_0$.  Using the parametrization   (see Fig.~\ref{fig:kinematic}) 
\begin{align}\label{eq:zwparam}
(z_1, z_ 2, z_3, z_4)  = (0  , \,   z  , \,  1   , \,  \frac{z-w }{1-w} )\,, 
\end{align} 
the master integrals evaluate to polylogarithmic functions up to weight three in $(z,\bar z)$ and $ (w, \bar w)$. 

The boundary integrals $B_{1-14}, C_0$ all arise from three-point topologies  $B^{[ijk]}_{1,0,1,1}$ with different detector orientations. 
For $(i,j,k)=(1,2,3)$, the master integrals define two dilogarithmic and two logarithmic functions, whose symbol alphabets read
\begin{align} 
\mathcal{A}_{123} =   \left\{ z,  \bar z , 1-z, 1-\bar z , 1- |z|^2 \right\} \,. \nn
\end{align} 

The type-I and II topologies contain 18 four-point master integrals $A_{1-18}$ which couple to four distinct boundary integral families:  $B^{[123]}_{1,0,1,1}$,  $B^{[124]}_{1,0,1,1}$, $B^{[134]}_{1,0,1,1}$ and $B^{[234]}_{1,0,1,1}$. 
Their symbol alphabets are drawn from the set 
\begin{align}
 \overline{\mathcal{A}}_{123} \cup  \overline{\mathcal{A}}_{124} \cup  \overline{\mathcal{A}}_{134} \cup  \overline{\mathcal{A}}_{234} 
  \cup \mathcal{A}_{\text{Quad}}\,,  \nn
\end{align}
where $\overline{\mathcal{A}}_{123} \equiv \mathcal{A}_{123} \cup \{z-\bar z \}  $ and 
\begin{align}
 \mathcal{A}_{\text{Quad}} & \equiv  \{   w \bar z - \bar w z , \,  w \bar z - \bar w , \,  z \bar w -  w ,   \nn   \\
& 1- w - \bar w +  z \bar w ,  \,  1- w- \bar w + w \bar z  \} \,.   \nn 
\end{align}
 $\overline{\mathcal{A}}_{124}, \overline{\mathcal{A}}_{134},\overline{\mathcal{A}}_{234}$ can be generated from $\overline{\mathcal{A}}_{123}$ through an $S_4$ symmetry transformation. 

\begin{figure}
\includegraphics[width=0.30\textwidth]{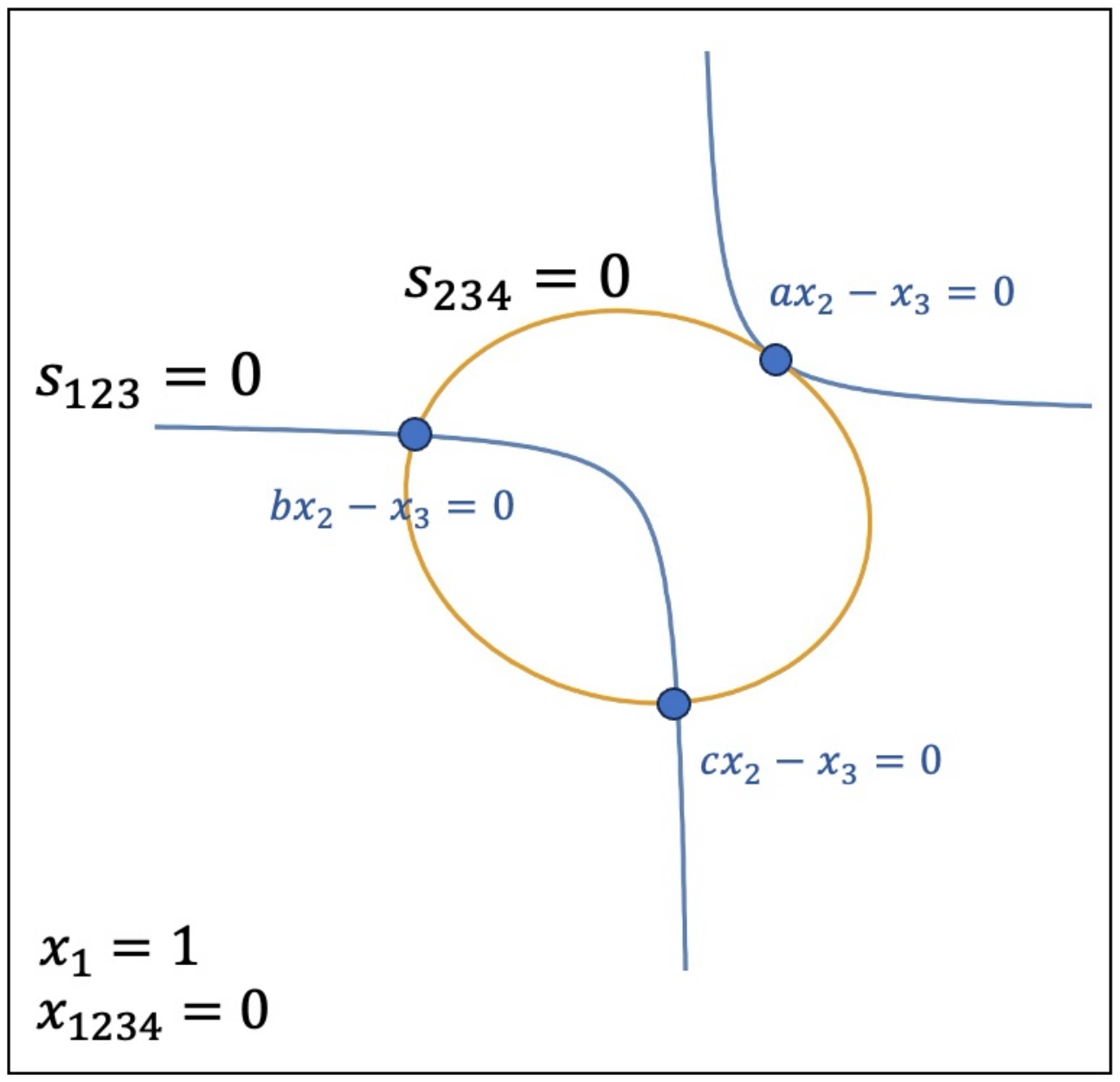} 
\caption{The intersection between the hyperplane $ x_{1234}=0$ and the hypersurfaces ${\blue s_{123}}=0$ and ${\blue s_{234}}=0$ projected onto the $(x_2, x_3)-$plane, gives cubic-root letters in the ${\rm{E^4C}}$.}
\label{fig_cubic}
\end{figure}

In the type-III topologies we identify 11 four-point master integrals which couple to two boundary integral families, $B^{[123]}_{1,0,1,1}$ and $B^{[234]}_{1,1,0,1}$, related by \emph{reflection}. 
The type-III integrands do not factorize linearly in terms of $(z, \bar z, w, \bar w)$. One must further introduce roots of a cubic polynomial, $(a,b,c)$, defined by 
\begin{align}\label{eq:abc_def}
& a\, b\, c = -|z|^2 |w|^2 , \quad a+ b+ c = 1- w - \bar w - z -\bar z\,, \notag \\ 
& a ^{-1}+b^{-1} +c^{-1}  = 1-z^{-1} - \bar{z}^{-1}- w^{-1} - \bar{w}^{-1} \,. 
\end{align}
The symbol alphabet of the type-III integrals are drawn from the set
\begin{align}
& \mathcal{A}_{123}  \cup \mathcal{A}_{\text{Tri}^2} \cup \{ 1 \leftrightarrow 4 ,    2 \leftrightarrow 3    \}  \nn   , 
\end{align}
where
\begin{align}\label{ATri2}
\mathcal{A}_{\text{Tri}^2}  =& \left\{|z|^2- |w|^2 ,  z- |w|^2 , \bar z - |w|^2\,,    \right.  \\
& \;  \left. 
   \frac{a}{b},  \frac{a+ |w|^2}{ b+ |w|^2} ,   \frac{a+ w}{ b+ w}   \frac{b+ \bar w}{ a + \bar w}   \right\}\cup \{a\rightarrow b, b\rightarrow c \}\,, \nn 
\end{align}
 and the \emph{reflection} maps $(z, w) $ to  $ (1/w, 1/z)$ and $(a,b,c) $ to $ (1/a,1/b,1/c)$. The  origin of the cubic-root letters can be understood from the maximal-cut condition ${\blue s_{123}} = { \blue s_{234} } = x_{1234}=0$,  whose solutions, when projected onto the $(x_2, x_3)-$plane, define a singular cubic curve $ (a\,  x_2 - x_3) (b \,x_2 - x_3)  (c\, x_2 - x_3) =0 $, as illustrated in Fig.~\ref{fig_cubic}.  The structure of the master integrals in this sector are discussed in App. A. Despite the cubic-root dependence,  the master integrals are single-valued functions whose branch cuts cancel on the Euclidean sheet \cite{Brown:2004ugm,Dixon:2012yy,Bourjaily:2022vti}. They all satisfy the expected first-entry condition \cite{Gaiotto:2011dt}, that the first entry of the symbol is $|z_{ij}|^2$. In App. B  we define the full set of master integrals $A_{1-28}$, $B_{1-14}$ and $C_0$, and provide their transformation into a basis of uniform transcendental weight. 
 

\subsection{Results}

Integrating the squared form factor, we obtain an analytic result for the collinear limit of the $\rm E^4C$ in $\mathcal{N}=4$ sYM. We find that the LO $\rm E^4C$ function space is comprised of polylogarithms up to weight 3, 
whose symbol alphabets are drawn from the set  
\begin{align}
\mathcal{A}_4 (1,2,3,4) & = \overline{\mathcal{A}}_{123} \nn 
 \cup \mathcal{A}_{\text{Quad}} \cup \mathcal{A}_{\text{Tri}^2}    \cup \text{perm} (1, 2, 3,  4)\,. \nn 
\end{align} 
Interestingly, the first entry of the symbol only sees five letters:  $\{ |z|^2, |1-z|^2, |w|^2, |1-w|^2, |z-w|^2 \} $. 
 The final answer can be expressed in terms of a basis of single-valued polylogarithmic functions as well as their images under $S_4$-permutations of the detector orientations, 
\begin{align} 
\hskip-2.5mm
{\rm{E^4C}}^{\rm LO, coll}_{\mathcal{N}=4} (z_{ij})& = \frac{ \sum\limits_{I=1}^{51} R_I(z,\omega)  \, F_I(z,\omega)}{|z_{12}|^2 |z_{23}|^2 |z_{34}|^2}   +  \text{perm}(1,2,3,4)\,. \nn  
\end{align}
%
%
%
%
%
%
In the ancillary files, we provide the analytic expressions for the basis $\vec{F}$ and algebraic coefficients $\vec{R}$. 

We have checked our result through direct numerical integration over the energy fractions. An additional check comes from the investigation of kinematic limits. In particular, we can access the triple-collinear limit by sending $z_1, z_2, z_3$ to the origin;  and the double-collinear limit by taking $|z_{12}|^2, |z_{34}|^2$ to zero simultaneously.  
In these limits the ${\rm E^4C}$ factorizes \cite{Chen:2019bpb}, 
\begin{align} 
&{\rm{E^4C}} \xrightarrow{z_1, z_2, z_3 \sim 0 } \, \frac{4}{ 3 |z_4|^2 } {\rm{E^3C}} (z_1, z_2, z_3)\,,  \nn  \\
&{\rm{E^4C}} \xrightarrow{z_{12}, z_{34} \sim 0 } \frac{2}{ 3 |z_{23}|^2 } \, {\rm{E^2C}} (z_1, z_2) \times {\rm{E^2C}} (z_3, z_4) \,, \nn 
\end{align}
into lower-point correlators, $\rm {E^2 C}, \rm {E^3 C}$, defined in Eq.~(\ref{NpointDef}). 
\subsection{Outlook}

Our result provides a number of directions for further exploring the perturbative structure of physical observables. First, it is important to study the singularity structure of the result to understand if its alphabet can be predicted a priori. There has been recent progress in the study of the Landau equations \cite{Mizera:2021icv,Fevola:2023fzn,Fevola:2023kaw}, intersection theory \cite{Mizera:2017rqa,Mizera:2019vvs,De:2023xue}, 
 as well as the structure of integrals directly in Feynman parameter space \cite{Arkani-Hamed:2017ahv,Gong:2022erh,Arkani-Hamed:2022cqe,Bourjaily:2020wvq,Britto:2023rig}, all of which are ideally suited for analyzing the integrals appearing in the ${\rm{E^\N C}}$.
Second, it is interesting to push towards higher point correlators, and to understand if statements can be made about the function space of the generic ${\rm{E^\N C}}$.
Third, our result suggests to further explore the origin and  the role of dual conformal symmetry for physical observables. Finally, it is highly desirable to extend our calculation to real world QCD, where our insight into the $\mathcal{N}=4$ sYM alphabet should prove invaluable.


\medskip

\noindent \textbf{Acknowledgements:}    
We acknowledge enlightening discussions with  Nima Arkani-Hamed, Hofie Hannesdottir, Gregory Korchemsky, Andrzej Pokraka, Gang Yang, Xiaoyuan Zhang and Hua Xing Zhu. 
This work is supported by the National Natural Science Foundation of China under Grant No. 12357077. KY would like to thank the sponsorship from Yangyang Development Fund. IM is supported by DOE-HEP-GR120647.





\appendix

\vspace{20pt}

\section{Appendix A}\label{sec:app_a}


In the type-III topologies,  the origin of the cubic-root letters  is sector $({\blue 2},{\blue 3},4)$, where  the solutions to 
  maximal-cut condition ${\blue s_{123}} = { \blue {\blue s_{1234}} } = x_{1234}=0$ 
  define  three projective lines.   Projected onto the $(x_2, x_3)-$plane, the line equations become a singular cubic curve $ (a\,  x_2 - x_3) (b \,x_2 - x_3)  (c\, x_2 - x_3) =0 $, with $a,b,c$ defined in Eq.~\eqref{eq:abc_def}. 
 Examining the {\it leading singularity} \cite{Cachazo:2008vp,Wasser:2022kwg,Henn:2020lye} of the master integrals in sector $({\blue 2},{\blue 3},4)$, we observe a subsystem $\{f_1, f_2 \} \equiv \frac{|z_{13}z_{34}|^2}{|z_{23}|^4} \{A_{22}, A_{23} \}$ whose leading singularities contain odd structures in the cubic roots, 
\begin{widetext}
\begin{align}\label{dlogf12}
f_1 &= 
\frac{1}{ (c-a)(a-b)} \int d \ln  x_3 \wedge d  \ln \frac{a \, x_2 -x_3 }{ b\,  x_2 - x_3}  \wedge  d \ln \frac{\blue D_2}{D_4}  \wedge   d \ln \frac{\blue D_3 }{D_4} 
 \nn \\
&
- \frac{1}{ (b-c)(c-a)} \int d \ln  x_3 \wedge d \ln \frac{b \, x_2 -x_3 }{ c\,  x_2 - x_3}  \wedge  d \ln \frac{\blue D_2}{D_4}  \wedge   d \ln \frac{\blue D_3 }{D_4}  \nn \\
f_2  &= \frac{a}{ (c-a)(a-b)} \int d \ln x_3 \wedge  \ln \frac{a \, x_2 -x_3 }{ b\,  x_2 - x_3}  \wedge  d \ln \frac{\blue D_2}{D_4}  \wedge   d \ln \frac{\blue D_3 }{D_4} 
 \nn \\
&
- \frac{c}{(b-c)(c-a)} \int d \ln  x_3  \wedge  \ln \frac{b \, x_2 -x_3 }{ c\,  x_2 - x_3}  \wedge  d \ln \frac{\blue D_2}{D_4}  \wedge   d \ln \frac{\blue D_3 }{D_4}\,.    
\end{align}
\end{widetext}
The above expressions have manifest cyclic symmetry. Thus we introduce pure functions $F_{a,b,c}$ through
\begin{align} 
f_1 & \equiv \frac{F_a}{(c-a)(a-b)} +   \frac{F_b}{(a-b)(b-c)} +  \frac{F_c}{(b-c)(c-a)} \,,   \quad \nn \\
f_2 & \equiv \frac{a\, F_a }{(c-a)(a-b)} + \frac{b\, F_b}{(a-b)(b-c)}  +\frac{c\, F_c}{(b-c)(c-a)}\,.  \nn
\end{align}
Eq.~\eqref{dlogf12} defines two dlog integrals $\{ F_a- F_b,  F_b- F_c \} $, whose symbol alphabets contain the odd cubic-root letters given in the last line of Eq.~\eqref{ATri2}. More specifically for $F_a$,  we  read off the following cubic-root letters in  its symbol: 
\begin{align} 
  & a \, ,  a+ |z|^2 \, , a+ |w|^2\, , u_a \equiv  (a+w)(a+\bar w) \, ,   \nn \\
  &  v_a \equiv (a+ z)(a+ \bar z) \,, \,    \frac{a+w}{a+\bar w} \, ,\,   \frac{a+z}{a+\bar z} \,.  \nn 
\end{align}
The identities 
$u_a = \frac{|w|^2 (a+|w|^2) (b+|z|^2)(c+|z|^2)}{ b\,c\,  ( |w|^2-|z|^2)} ,
v_a = \frac{ a\, (a+|z|^2) (b+|w|^2)(c+|w|^2)}{ |w|^2 ( |w|^2-|z|^2)} $,
together with the shuffle algebra \cite{Ree:1958},  allow us to show 
in the first and second entry of  $\mathcal{S}[F_a]$, the cubic-root alphabets appear as cyclic products: $ a\, b\, c\,,  (a+|z|^2) (b+|z|^2) (c+|z|^2)$ and $ (a+|w|^2) (b+|w|^2) (c+|w|^2) $. 
Therefore $\mathcal{S}[F_a]$ contains only rational letters in the first and second entry, and in particular the first-entry condition  for single-valuedness is satisfied.   Moreover, in the cyclic combination $\{ F_a+ F_b+F_c\} $, the dependence on cubic roots fully cancels, and its symbol letters are rational in all entries.

The top sector $({\blue 2},{\blue 3},4,5) $ contains two additional finite dlog integrals $f_3, f_4$, which are linear combinations of $A_{23}, A_{24},A_{25},A_{26}$. They can be cast into the following dlog form, 
\begin{widetext}
\begin{align}
f_{3} &= \int d\ln x_3 \wedge  d \ln \frac{w \, x_2 +x_3}{\bar w \, x_2 +x_3}  \wedge  d \ln \frac{\blue D_2}{D_4}  \wedge   d \ln \frac{\blue D_3 }{D_5} \,,   \nn \\
f_4 &=\frac16 \int d\ln  x_3  \wedge  d \ln \frac{[ (a \, x_2 -x_3) (b \, x_2 -x_3) (c\,  x_2 -x_3)]^2}{[(w \, x_2 +x_3)(\bar w \, x_2 +x_3)]^3}  \wedge  d \ln \frac{\blue D_2}{D_4}  \wedge   d \ln \frac{\blue D_3 }{D_4}    \nn \\
& +\frac12  \int d\ln  x_3  \wedge  d \ln [( w \, x_2 +x_3)(\bar w \, x_2 +x_3) ]\wedge  d \ln \frac{\blue D_2}{D_4}  \wedge   d \ln \frac{ D_5 }{D_4}\,, \nn 
\end{align}
\end{widetext}
which suggests that $f_3,f_4$ do not depend on the cubic roots.  While the first piece of $f_4$ contains  $F_a+ F_b+F_c$, its dependence on the cubic roots is spurious.

\section{Appendix B}\label{sec:app_b}

We provide the definition of the master integrals in the Type-I, II  and III subtopologies. The integrands for the four-point master integrals $A_{1-28}$ are  
\begin{align}
& A_{1,2} :    \quad   \frac{x_3  |z_{12}|^2 }{  {\blue s_{123}} \, x_{34}^2 \, x_{1234}}  \,,   \quad 
 \frac{|z_{12}|^2 }{  {\blue s_{123}}\, x_{34} \, x_{1234}}   , \quad 
\nn\\
& 
A_3 :  \quad    \frac{ |z_{12}|^2 }{   {\blue s_{123}}\, x_{234} \, x_{1234}}\,,    \nn  \\
& A_{4,5} : \quad  \frac{ x_3 |z_{23}|^2   }{   {\blue s_{123}}\, x_{123}\, x_{34} \, x_{1234}}   , \quad   \frac{ x_2   |z_{23}|^2  }{   {\blue s_{123}}\, x_{123}\, x_{34} \, x_{1234}} \,,     
\nn \\ 
& A_{6,7} :  \quad   \frac{ x_3   |z_{23}|^2  }{   {\blue s_{123}}\, x_{123}\, x_{234} \, x_{1234}}    , \quad  \frac{ x_1  |z_{23}|^2  }{   {\blue s_{123}}\, x_{123}\, x_{234} \, x_{1234}} .      
\nn \\
 & A_8 : \quad   \frac{x_3 |z_{12}|^4  }{  {\blue s_{1234}}^2\, x_{34}  }   , \quad    
A_9 : \quad     \frac{ |z_{12}|^2  }{  {\blue s_{1234}}\, x_{34} \, x_{1234}}  \,,   \quad
  \nn \\
&
A_{10} : \quad     \frac{ |z_{12}|^2}{   {\blue s_{1234}}\, x_{234} \, x_{1234}}\,,   \quad
A_{11} : \quad \frac{ x_3   |z_{12}|^4  }{   {\blue s_{1234}} \, {\blue s_{123}}\, x_{34} }   , \quad 
\nn\\
&    
A_{12,13} : \quad  \frac{ x_2 |z_{12}|^4   }{   {\blue s_{1234}}\, {\blue s_{123}}\, x_{234} } \,,   \quad 
  \frac{ x_3 |z_{12}|^4  }{   {\blue s_{1234}}\, {\blue s_{123}}\, x_{234} } \,,   \nn \\ 
& A_{14-16} :  
  \nn \\
  & \qquad 
   \frac{ x_1 |z_{12}|^4  }{   {\blue s_{1234}}\, {\blue s_{123}}\, x_{1234} }    , \quad 
   \frac{ x_2  |z_{12}|^4 }{   {\blue s_{1234}}\, {\blue s_{123}}\, x_{1234} }   ,   
 \quad  \frac{ x_3 |z_{12}|^4}{   {\blue s_{1234}}\, {\blue s_{123}}\, x_{1234} }    ,     \nn \\
 & A_{17} :   \quad  \frac{ x_1 x_2 |z_{12}|^4  }{   {\blue s_{1234}}\, {\blue s_{123}}\, x_{34} \, x_{1234} }    , \quad 
 A_{18} : \quad    \frac{ x_2^2 |z_{12}|^4 }{   {\blue s_{1234}}\, {\blue s_{123}}\,  x_{234} \, x_{1234} }    .    
 \nn \\
 & A_{19-23} :   \quad 
 \nn\\
 & \qquad   \frac{ x_1 x_2 |z_{23}|^4 }{ {\blue s_{123}}\, {\blue s_{234}}\, x_{1234}^2  }  , \quad     
    \frac{ x_1 x_3  |z_{23}|^4 }{  {\blue s_{123}}\, {\blue s_{234}} \, x_{1234}^2}  \,,   \quad   
    \frac{ x_2 x_4  |z_{23}|^4  }{   {\blue s_{123}}\, {\blue s_{234}} \, x_{1234}^2 }\,,    \nn  \\
& \qquad  
   \frac{ x_2 |z_{23}|^4 }{   {\blue s_{123}} \, {\blue s_{234}}\, x_{1234} }   , \quad    
    \frac{ x_3 |z_{23}|^4  }{   {\blue s_{123}}\, {\blue s_{234}}\,  x_{1234} } \,,   \quad  \nn \\
& A_{24,25} : \quad    \frac{ x_2^2   |z_{23}|^4}{   {\blue s_{123}}\, {\blue s_{234}}\, x_{234}\, x_{1234} }    , \quad 
  \frac{ x_2 x_3   |z_{23}|^4  }{   {\blue s_{123}}\, {\blue s_{234}}\,x_{234} \,x_{1234} }       ,     \nn \\
 & A_{26- 28} :  \quad
 \nn \\
&  \hspace{-2.0mm} \frac{ x_2 x_3 |z_{23} z_{14}|^2    }{   {\blue s_{123}} {\blue s_{234}} x_{123} x_{234} } ,   \quad   \frac{ x_2^2  |z_{23} z_{14}|^2  }{   {\blue s_{123}}{\blue s_{234}}x_{123}  x_{234} } ,      \quad 
 \frac{ x_3^2 |z_{23} z_{14}|^2  }{   {\blue s_{123}}{\blue s_{234}}x_{123}  x_{234} }\,. \nn
\end{align}
The integrands for the three-point boundary integrals $B_{1-14}$ are 
\begin{align}
&B_{1-3} : \quad 
   \frac{x_{1}|z_{12}|^2}{{\blue s_{123}} \,  x_{123}^2}, \quad  \frac{x_{2}  |z_{12}|^2  }{{\blue s_{123}} \, x_{123}^2} , \quad   \frac{|z_{12}|^2}{{\blue s_{123}}  \, x_{123}}, \quad
\nn \\
&
 B_4 :  \quad  
   \frac{x_2  |z_{12}|^2 }{{\blue s_{123}}  \, x_{23} \, x_{123}} , \nn \\
&B_{5,6} : \quad  \frac{x_{1}  |z_{12}|^2    }{{ \blue s_{124} } \,  x_{124}^2}, \quad  \frac{|z_{12}|^2  }{{\blue s_{124}}  \, x_{124}}, \quad B_7 : \quad   \frac{x_2  |z_{12}|^2 }{{\blue s_{124}}  \, x_{24} \, x_{124}},  \nn \\
&B_{8} : \quad     \frac{|z_{34}|^2 }{{\blue s_{134}}  \, x_{134}}, \quad B_{9} : \quad  \frac{x_3   |z_{34}|^2  }{ {\blue s_{134}}  \, x_{24} \, x_{134}},  \nn \\
&B_{10-12} : \quad     \frac{x_{2} |z_{23}|^2}{{\blue s_{234}} \,  x_{234}^2}, \quad   \frac{x_{3}|z_{23}|^2 }{{\blue s_{234}} \, x_{234}^2} , \quad  \frac{|z_{23}|^2 }{{\blue s_{234}}  \, x_{234}}, \quad  \nn \\
& B_{13} : \quad   \frac{x_3  |z_{23}|^2 }{{\blue s_{234}}  \, x_{34} \, x_{234}} ,  \quad
\nn \\
& 
B_{14}: \quad  \frac{x_3  |z_{23}|^2 }{{\blue s_{234}}  \, x_{23} \, x_{234}} \,.  \nn 
\end{align}
The two-point boundary integrals reduce to a constant function, which we define as $C_0 =1$.
The complete basis of master integrals $\{ \vec{A}, \vec{B}, C_0 \}$ can be brought into  a UT basis $\{\vec{I}_A,  \vec{I}_B, C_0 \}$ by a linear transformation. We provide the $43 \times 43$ transformation matrix $T$ in the ancillary files.

\bibliography{joh_more_refs}{}
\bibliographystyle{apsrev4-1}

\end{document}